\documentclass{aa}
\usepackage{graphicx}
\usepackage[varg]{txfonts}
\usepackage{cases}
\usepackage{natbib}

\begin{document}
   \title{Remote coronal dimmings related to a circular-ribbon flare}

   \author{Q. M. Zhang\inst{1,2,3} and R. S. Zheng\inst{4}}

   \institute{Key Laboratory for Dark Matter and Space Science, Purple
              Mountain Observatory, CAS, Nanjing 210033, PR China \\
              \email{zhangqm@pmo.ac.cn}
              \and
              State Key Laboratory of Lunar and Planetary Sciences, Macau University of Science and Technology, Macau, China \\
              \and
              State Key Laboratory of Space Weather, Chinese Academy of Sciences, Beijing 100190, PR China \\
              \and
              Shandong Provincial Key Laboratory of Optical Astronomy and Solar-Terrestrial Environment, Weihai 264209, PR China \\
              \email{ruishengzheng@sdu.edu.cn}
              }

   \date{Received; accepted}
    \titlerunning{Remote coronal dimmings}
    \authorrunning{Zhang \& Zheng}

 \abstract
   {}
   {In this paper, multiwavelength observations of remote coronal dimmings related to an M1.1 circular-ribbon flare (CRF) in active region (AR) 12434 are reported.}
   {The confined flare without a CME was observed by AIA and HMI on board SDO on 2015 October 16.
   Global three-dimensional (3D) magnetic fields before flare were obtained using the potential field source surface modeling.}
   {A few minutes before the flare hard X-ray peak time (06:13:48 UT), small-scale, weak dimming appeared $\sim$240$\arcsec$ away from the flare site, 
   which can be observed by AIA only in 131 and 171 {\AA}. 
   Afterwards, long and narrow dimmings became evident in all AIA EUV passbands except 304 {\AA}, 
   while localized core dimming was not clearly observed near the flare site.
   The large-area dimmings extended southeastward and the areas increased gradually. 
   The total area of dimmings reaches (1.2$\pm0.4$)$\times$10$^4$ Mm$^2$ in 193 {\AA}. 
   The maximal relative intensity decreases in 171 and 193 {\AA} reach 90\% and 80\%, respectively. 
   Subsequently, the dimmings began to replenish and the area decreased slowly, lasting for $\geq$3 hr. 
   The remote dimmings and AR 12434 were connected by large-scale coronal loops. 
   The remote dimmings were associated with the southwest footpoints of coronal loops with weak negative polarities.
   Possible origins of remote dimmings are discussed.}
   {}


 \maketitle

\section{Introduction} \label{s-intro}
Coronal dimmings are transient regions of greatly reduced emission in extreme-ultravoilet (EUV) and soft X-ray (SXR) wavelengths \citep{hud96,thom98,thom00}.
They are tightly related to CMEs and/or solar flares \citep{rein09,kri17,diss18a,diss18b,diss19,ver19}. 
When a CME rises and erupts into the interplanetary space, the coronal source region is impulsively evacuated and the plasma density is heavily depleted \citep{har00,har03,mas14}. 
The percentage of density depletion could reach up to 50\%$-$70\% \citep{vann18}.
Persistent upflows at speeds of hundreds of km s$^{-1}$ have been observed using the spectroscopic observations \citep{harr01,harr07,att10,doll11,tian12,ver19}.
The Doppler velocities depend on the formation temperatures of ions \citep{jin09}, and the most significant dimmings can be observed at temperatures of a few MK \citep{zhu04,robb10}.

According to the location, coronal dimmings can be classified into deep core dimmings near the flare site and widespread diffuse dimmings \citep{zar99,man05,att07,del07}. 
The deep core dimmings or twin dimmings are related to the footpoints of a flux rope, while the secondary diffuse dimmings result from continuing field line 
stretching during the eruption \citep{man07,mik11,diss18b,vann18}. The widths of dimmings are positively correlated with the CME widths \citep{rein08}. 
The velocity and mass of CMEs are found to be intimately related to the slope and depth of the dark voids left behind CMEs \citep{mas16}. 

In most cases, coronal dimmings appear in the impulsive phase of a flare that is consistent with the acceleration phase of the corresponding CME \citep{cheng16}.
The typical evolution of dimming area is characterized by a sharp rise followed by a slow recovery via magnetic reconnection \citep{att06,rein08}.
The total area and unsigned magnetic flux of dimmings are $\sim$10$^{10}$ km$^2$ and $\sim$10$^{21}$ Mx \citep{diss18b}.
The total lifetime of dimmings ranges from 3 to 12 hr with an average value of $\sim$8 hr \citep{rein08}.
Using the high-resolution and multiwavelength observations from the Atmospheric Imaging Assembly \citep[AIA;][]{lem12} on board the Solar Dynamics Observatory (SDO), 
\citet{zqm17} discovered pre-flare coronal dimmings, which lasted for $\sim$96 minutes before the impulsive phase of an eruptive flare. 
It is proposed that pre-flare dimming serves as a precursor of twin dimmings. Additional events about pre-eruption dimmings were investigated and reported \citep{qiu17,wang19}.

During the whole evolution of NOAA active region (AR) 12434 in 2015 October, a series of homologous and short-lived circular-ribbon flares (CRFs) were produced \citep{zqm16a,zqm16b}.
\citet{zqm19} calculated the energy partition of two M1.1 CRFs taking place on October 15 and 16, respectively. 
In this paper, remote coronal dimmings related to the M1.1 flare on October 16 are investigated to explore the relationship between CRFs and coronal dimmings.
Observations and data analysis are described in Sect.~\ref{s-data}. Discussions and a brief summary are shown in Sect.~\ref{s-disc}.

\section{Observations and data analysis} \label{s-data}
The M1.1 flare was observed by AIA \citep{zqm19}. AIA takes full-disk images in 7 EUV wavelengths (94, 131, 171, 193, 211, 304, and 335 {\AA}) 
with a cadence of 12 s and in two UV wavelengths (1600 and 1700 {\AA}) with a cadence of 24 s.
The photospheric line-of-sight (LOS) magnetograms are observed by the Helioseismic and Magnetic Imager \citep[HMI;][]{sch12} on board SDO with a cadence of 45 s.
The spatial resolutions of AIA and HMI are 2\farcs4.

\begin{figure}
\includegraphics[width=8cm,clip=]{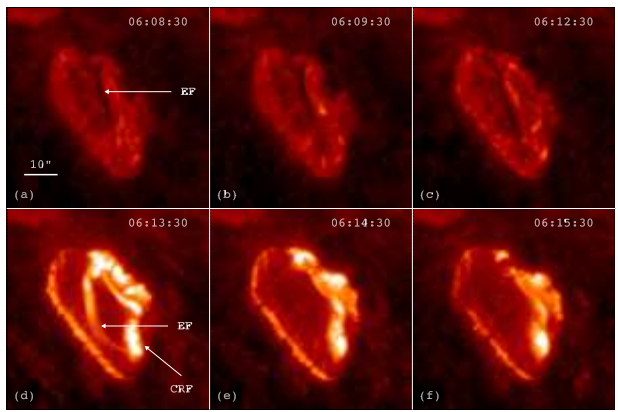}
\centering
\caption{Six snapshots of the original AIA 304 {\AA} images. The white arrows point to the eruptive filament (EF) that triggered the circular-ribbon flare (CRF).
The field of view is 60$\arcsec$$\times$60$\arcsec$. The evolution of EF is shown in a movie (\textit{anim0.mov}) available in the online edition.}
\label{fig1}
\end{figure}

\begin{figure}
\includegraphics[width=8cm,clip=]{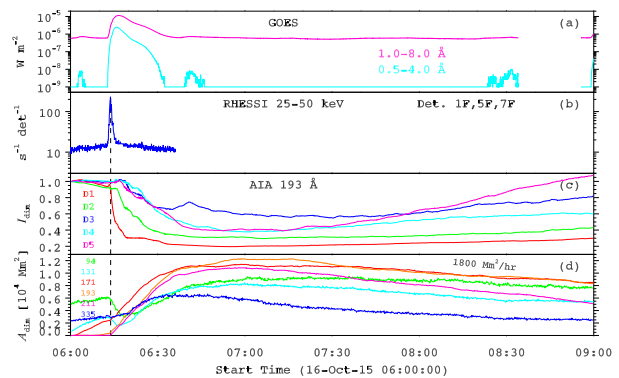}
\centering
\caption{(a) GOES light curves of the flare. (b) HXR light curve at 25$-$50 keV.
(c) Time evolutions of the average intensities of ``D1"-``D5" in Fig.~\ref{fig4}(e).
(d) Time evolutions of the total area of remote dimmings in different wavelengths.
The black dashed line denotes the time at 06:13:48 UT.}
\label{fig2}
\end{figure}

Figure~\ref{fig1} shows six snapshots of the AIA 304 {\AA} images. It is clear that the CRF was triggered by the eruption of a mini-filament at $\sim$06:13 UT.
During its eruption, the arc-shaped eruptive filament (EF) was heated and became bright in 304 {\AA}.
SXR light curves of the flare with a cadence of $\sim$2.05 s are plotted in Fig.~\ref{fig2}(a). The flux in 1$-$8 {\AA} increased rapidly from $\sim$06:11 UT to $\sim$06:16 UT, 
which was followed by a gradual decay phase until $\sim$06:35 UT. 
The hard X-ray (HXR) flux at 25$-$50 keV observed by the Reuven Ramaty High-Energy Solar Spectroscopic Imager \citep[RHESSI;][]{lin02} is plotted in Fig.~\ref{fig2}(b).
The peak time (06:13:48 UT) indicative of the highest energy release rate is denoted by the black dashed line.

\begin{figure}
\includegraphics[width=8cm,clip=]{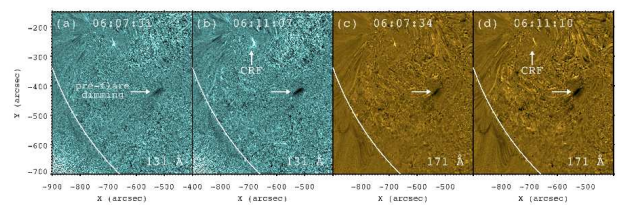}
\centering
\caption{AIA base-ratio images in 131 and 171 {\AA}.
The horizontal and vertical arrows point to the pre-flare dimming (dark area) and CRF. 
The evolution of pre-flare dimming is shown in a movie (\textit{anim1.mov}) available in the online edition.}
\label{fig3}
\end{figure}

The confined flare was neither associated with a CME nor with localized core dimming near the flare site. 
Instead, it was related to remote dimmings $\sim$240$\arcsec$ away in the quiet region. 
To better illustrate the coronal dimmings, the base-ratio technique is applied by taking the EUV images at $\sim$06:00:00 UT as base images \citep{diss18a}.
Figure~\ref{fig3} shows 4 snapshots of base-ratio images during 06:07$-$06:11 UT in 131 {\AA} and 171 {\AA}.
It is surprising that small-scale, narrow dimming appeared to the southwest of flare a few minutes before the onset of flare, which is considered as pre-flare dimming.
The area increased very slowly with the intensity decreasing.
We have checked the base-ratio images in other EUV wavelengths and found that pre-flare dimming was prominent only in 131 and 171 {\AA} 
with formation temperatures of $\log T\approx5.6-5.8$ in the lower corona.

\begin{figure}
\includegraphics[width=8cm,clip=]{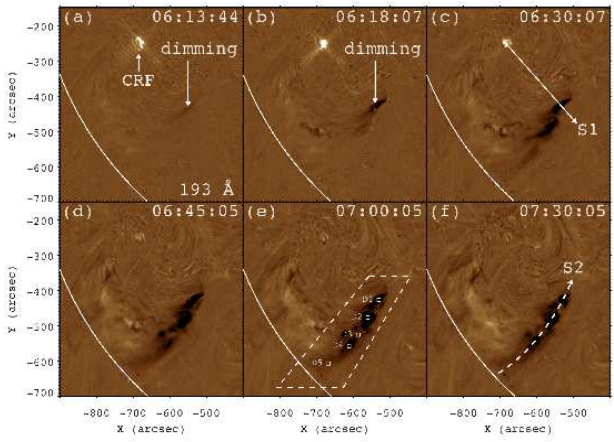}
\centering
\caption{AIA base-ratio images in 193 {\AA} showing the remote coronal dimmings (dark area). 
The evolution of large-area dimmings is shown in a movie (\textit{anim2.mov}) available in the online edition.}
\label{fig4}
\end{figure}

In Fig.~\ref{fig4}, the 193 {\AA} base-ratio images illustrate the evolution of large-area remote dimmings after the onset of flare. 
Long and narrow dimmings, which were cospatial with the pre-flare dimming, appeared shortly after $\sim$06:14 UT. 
As time went on, the dimmings extended southeastward with the area increasing rapidly. The dimmings were replenished and the area reduced slowly after the maxima.
The temporal and spatial relationships between pre-flare dimming and large-area dimmings suggest that remote dimmings started from the pre-flare dimming and 
developed into large-area dimmings. Hence, the pre-flare dimming may serve as a precursor of large-area dimmings.

In Fig.~\ref{fig4}(e), five small boxes (``D1''-``D5'') within the dimmings are selected and investigated.
Time evolutions of the average base-ratio intensities within the five boxes in 193 {\AA} are plotted in Fig.~\ref{fig2}(c). The intensity evolution is characterized by a sharp
decrease followed by a gradual recovery. It is obvious that the intensity depletion of D1 preceded that of D2.
The depletions of D3-D5 were further delayed, which is consistent with the southeastward extension of large-area remote dimmings. 
Moreover, the relative intensity decrease of D1 reaches $\sim$80\%, which is considerably higher than that of D3 ($\sim$45\%).
Owing to the different formation temperatures and heights of AIA EUV passbands, the degrees of intensity depletion are different accordingly. 
In Table~\ref{table:1}, the maximal relative intensity decreases of the large-area dimmings are listed. 
It is revealed that the dimmings can remarkably be detected in 171, 193, and 211 {\AA}, and are less remarkable in 94, 131, and 335 {\AA} \citep{vann18}. 

\begin{table}
\caption{Maximal area ($\times10^4$ Mm$^2$) and relative intensity decrease (\%) of the remote dimmings in different wavelengths.} 
\label{table:1}
\centering
\begin{tabular}{c|ccc}
\hline\hline
$\lambda$ ({\AA}) & 94 & 131 & 171 \\
\hline
Area & 0.67$-$1.28 & 0.4$-$1.42 & 0.78$-$1.70 \\
Percent & 15$-$70 & 20$-$75 & 55$-$92 \\
\hline
$\lambda$ ({\AA}) & 193 & 211 & 335 \\
\hline
Area & 0.84$-$1.79 & 0.70$-$1.78 & 0.28$-$1.29 \\
Percent & 45$-$80 & 30$-$55 & 10$-$55 \\
\hline
\end{tabular}
\end{table}

In Fig.~\ref{fig4}(e), the remote dimmings are encompassed by white dashed lines. The total area of dimmings is defined as the sum of pixels whose intensities 
are less than 85\% of the pre-flare level at 06:00 UT. Time evolutions of the dimming area in different wavelengths are plotted in Fig.~\ref{fig2}(d).
It is found that the area increased quickly to the apex around 07:16 UT before declining tardily. 
The lifetime of dimmings exceeded 3 hr, with a rise time of $\sim$1 hr and a recovery time of $\geq$2 hr. In Fig.~\ref{fig2}(d), the recovery rate ($\sim$1800 Mm$^2$ hr$^{-1}$) of large-area dimmings is labeled.
Assuming a constant recovery rate, the dimmings would be totally recovered $\sim$4.8 hr after 09:00 UT. Therefore, an upper limit of the lifetime of dimmings is nearly 8 hr.
The dimming area reaches $\sim$1.2$\times$10$^4$ Mm$^2$ in 193 {\AA} and amounts to only 6000$-$8000 Mm$^2$ in 131 and 335 {\AA}. 
In Table~\ref{table:1}, the maximal total areas of the remote dimmings in different wavelengths are listed when the intensity threshold changes from 75\% to 95\% of the pre-flare levels.

\begin{figure}
\includegraphics[width=8cm,clip=]{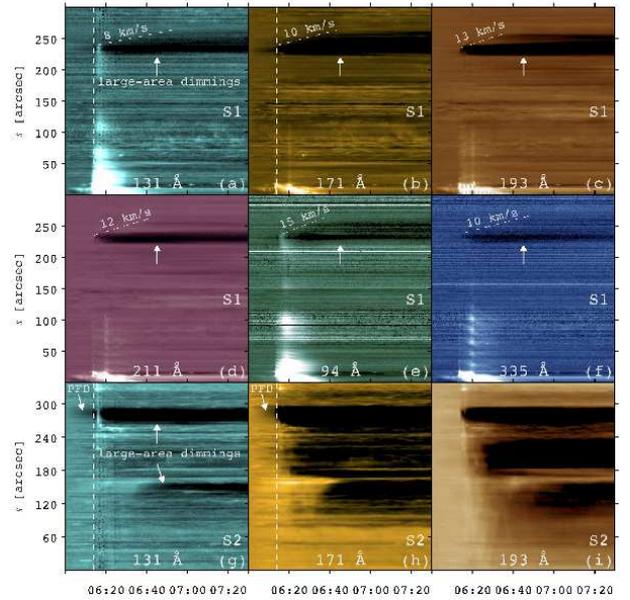}
\centering
\caption{(a)-(f) Time-distance diagrams of S1. In panels (a) and (b), the white dashed lines denote the HXR peak time (06:13:48 UT).
(g)-(i) Time-distance diagrams of S2.}
\label{fig5}
\end{figure}

In Fig.~\ref{fig4}(c), an artificial slice (S1) with a length of 300$\arcsec$ is selected, which starts from the flare core and goes across the large-area dimmings.
Time-distance diagrams of S1 in different wavelengths are displayed in Fig.~\ref{fig5}(a)-(f). It is obvious that localized core dimming near the flare was not observed. 
As soon as the flare occurred, remote dimmings appeared about 240$\arcsec$ away. 
The apparent speeds of expansion of dimmings along S1 are measured to be 8$-$15 km s$^{-1}$, with a mean value of $\sim$11 km s$^{-1}$.
In Fig.~\ref{fig4}(f), a curved slice (S2) with a length of 338$\farcs$5 is selected, which goes through the entire remote dimmings. 
Time-distance diagrams of S2 in different wavelengths are shown in Fig.~\ref{fig5}(g)-(i), indicating that the extraordinary large-area dimmings extended in the southeast direction.

\begin{figure}
\includegraphics[width=8cm,clip=]{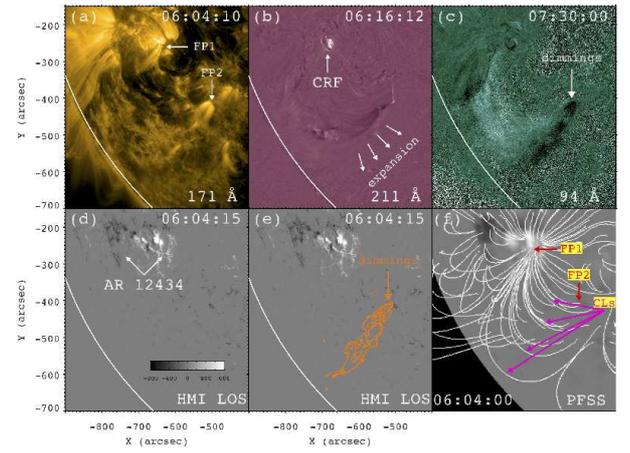}
\centering
\caption{(a) Original AIA 171 {\AA} image. The white arrows point to the footpoints of the large-scale, overlying CLs.
(b)-(c) Base-ratio images in 211 {\AA} and 94 {\AA}.
(d)-(e) LOS magnetogram, which is superposed by intensity contours of 193 {\AA} image in Fig.~\ref{fig4}(e).
(f) Global magnetic fields obtained from the PFSS modeling.
The evolution of remote coronal dimmings in 211 {\AA} and 94 {\AA} is shown in a movie (\textit{anim3.mov}) available in the online edition.}
\label{fig6}
\end{figure}

In Fig.~\ref{fig6}, the original AIA image in 171 {\AA} before flare is displayed in panel (a). Two arrows point to the footpoints (FP1 and FP2) of imaginary large-scale coronal loops (CLs),
which are more distinct in 211 {\AA}. In panel (b), the base-ratio image in 211 {\AA} at 06:16:12 UT is shown. The oblique arrows indicate the directions of quick expansion of the CLs. 
The apparent speeds of southwestward expansion are indicated in Fig.~\ref{fig5}.
In Fig.~\ref{fig6}(e), the LOS magnetogram at 06:04:15 UT is superposed by intensity contours of 193 {\AA} image in Fig.~\ref{fig4}(e).
FP1 in AR 12434 is associated with strong positive polarity, while FP2 in the quiet region is associated with weak negative polarities.
In Fig.~\ref{fig6}(f), the global magnetic fields obtained from the potential field source surface \citep[PFSS;][]{sch03} modeling
at 06:04:00 UT is depicted, where a series of closed magnetic loops are indicated by magenta arrows.
Figure~\ref{fig6}(c) shows the base-ratio image in 94 {\AA} at 07:30:00 UT, where the bright patch corresponds to large-scale CLs and the dark dimmings correspond to the footpoints (FP2) of CLs.
Combining the EUV observations, LOS magnetogram, and 3D magnetic configuration, it is concluded that FP1 and FP2 are the footpoints of large-scale, overlying CLs.

\section{Discussion and summary} \label{s-disc}
As mentioned in Sect.~\ref{s-intro}, there are two types of coronal dimmings in EUV wavelengths, 
core dimming and widespread diffuse dimmings that cover a large fraction of the solar surface \citep[e.g.,][]{thom98,del07,diss18b}.
They are mainly due to sudden density depletion when CMEs and/or flares take place \citep{har00,har03}.
After investigating three flare-related EIT waves, \citet{del00} found multiple dimmings: a first dimming between the bright front and the flare, 
a second dimming between the bright front and a magnetic dipole, and a third dimming in transequatorial loops connecting the flare region with another dipole.
In this work, the primary dimming surrounding the flare was not obvious. Only remote dimmings appeared directly far from the flare site in the quiet region.
The presence of remote dimmings in all EUV wavelengths except 304 {\AA} suggests that the dimmings are interpreted by density depletion rather than temperature variation \citep{mas14}.
The degrees of intensity decrease are comparable to those of CME-associated dimmings \citep{diss18b}.
The evolution of remote dimmings is characterized by a rapid increase followed by a gradual decay, which is similar to typical CME-associated dimmings \citep{rein08}.
Both the rise and recovery times are close to those of CME-associated dimmings \citep{rein08}. 
The total areas have the same order of magnitude as the CME-associated dimmings and are close to the median value (2.15$\times$10$^4$ Mm$^2$) \citep{diss18b}.

After studying two confined flares triggered by the filament eruption, \citet{yang18} discovered that the erupting filaments are guided by large-scale field lines and finally 
reach the remote site (see their Fig. 3). In a follow-up work, \citet{yang19} proposed a two-step evolution of a magnetic flux rope, the standard rising phase producing a pair of 
parallel ribbons and the external reconnecting phase, during which the eruptive flux rope reconnects with the large-scale, overlying CLs (see their Fig. 7). 
The external reconnection results in transfer and redistribution of the magnetic helicity. Meanwhile, the large-scale loops are heated and filled with hot plasma. 
A pair of newly formed ribbons are generated in the chromosphere.

In this study, the CRF was also triggered by an eruptive mini-filament.
During its initial rising phase, the filament pushed the CLs aside and resulted in small-scale, pre-flare dimming, which could be observed only in 131 and 171 {\AA} 
with formation temperatures of $\log T\approx5.6-5.8$ in the lower corona.
After the onset of CRF, the filament continued to ascend and dilate. However, the confinement from the CLs was so strong that the filament failed to evolve into a CME \citep{sun15}. 
Magnetic reconnection between the EF and CLs may also heat the CLs to high temperature observed in 94 {\AA}. 
Meanwhile, the CLs were continuously being pushed aside and underwent rapid expansion in the southwest direction. 
Therefore, the large-area remote dimmings, which were associated with the negative footpoints (FP2) of CLs, may result from density depletion of the CLs during their expansion, 
which is consistent with the fact that large-area dimmings were observed in all EUV passbands except 304 {\AA}.

In this work, we investigated the confined M1.1 CRF triggered by the eruption of a mini-filament without a CME on 2015 October 16.
A few minutes before the flare HXR peak time, small-scale, weak dimming appeared $\sim$240$\arcsec$ away from the flare.
Such pre-flare dimming was observed only in 131 and 171 {\AA}.
Afterwards, long and narrow dimmings became evident in all AIA EUV wavelengths except 304 {\AA}, while localized core dimming was absent near the flare site. 
The large-area dimmings extended southeastward and the areas increased gradually. 
The total area of remote dimmings reaches (1.2$\pm0.4$)$\times$10$^4$ Mm$^2$ in 193 {\AA}. The maximal relative intensity decreases in 171 and 193 {\AA} reach 90\% and 80\%, respectively. 
Subsequently, the dimmings began to replenish and the area decreased slowly, lasting for $\geq$3 hr.
The remote dimmings and AR 12434 were connected by large-scale, overlying CLs. 
The dimmings were associated with the southwest footpoints of CLs with weak negative polarities.
We propose that the remote coronal dimmings were probably created by density depletion of the large-scale CLs.

\begin{acknowledgements} 
The authors are grateful for the referee for valuable suggestions to improve the quality of this article.
We thank Y. N. Su, T. H. Zhou, L. Lu, and D. Li in Purple Mountain Observatory for their kind help. 
SDO is a mission of NASA\rq{}s Living With a Star Program. AIA and HMI data are courtesy of the NASA/SDO science teams. 
This work is funded by the NSFC grants (No. 11773079, 11790302, 11603013, U1731101), 
the Science and Technology Development Fund of Macau (275/2017/A),
the International Cooperation and Interchange Program (11961131002), the Youth Innovation Promotion Association CAS, 
and the project supported by the Specialized Research Fund for State Key Laboratories.
\end{acknowledgements}

\end{document}